\documentclass[fleqn,12pt]{wlscirep}
\usepackage{amsmath}
\usepackage{graphicx}
\usepackage{footmisc}
\usepackage{color}

\begin{document}
\title {Electron liquid state in the spin-$\frac{1}{2}$ anisotropic Kondo lattice
}
\author[1]{Igor N. Karnaukhov}
\affil[1]{G.V. Kurdyumov Institute for Metal Physics, 36 Vernadsky Boulevard, 03142 Kyiv, Ukraine}
\affil[*]{karnaui@yahoo.com}
\begin{abstract}
In the framework of the mean field approach, we provide analytical and numerical solution of the spin-$\frac{1}{2}$ anisotropic Kondo lattice for arbitrary dimension at half filling. Nontrivial solution for the amplitude of the field opens a gap in the fermion spectrum of an electron liquid in which local moments on the lattice sites are realized. The ground state in the insulator state is determined by a static $\mathbb{Z}_2$ field of local moments, which forms the lattice with a double cell, conduction electrons move in this field.  Due to hybridization between electron states a large Fermi surface is formed in the Kondo lattice. A gap in the quasi-particle spectrum is calculated depending on the magnitudes of exchange integrals for the  simple lattices with different dimension. The proposed approach is also valid for describing the Kondo lattice with weak anisotropy of the exchange interaction, which makes it possible to study the behavior of the spin-$\frac{1}{2}$ Kondo lattice with an isotropic exchange interaction.
\end{abstract}
\maketitle

\section*{Introduction}

The study of behavior of the Kondo-lattice is an actual problem of condensed matter physics.
Until now, the phase state of the Kondo lattice is an open page: we cannot understand the physical nature of the state of the Kondo insulator and cannot calculate the thermodynamic characteristics in this phase state. It is clear that, first of all, it is necessary to take into account the processes of scattering of band electrons with spin flip on local moments located at lattice sites. Thus, in the Kondo problem, these scattering processes lead to the Abrikosov-Suhl resonance \cite {W1,W2}. Simplified modifications of the Hamiltonian of the Kondo lattice  \cite{1,K0,KS} do not take these processes into account and, as a result, do not adequate describe the phase diagram  of the Kondo lattice.
In the numerical calculations of Anderson and Kondo lattices, small clusters are taken into account, which significantly affects the accuracy and reliability of the results of calculations \cite{A1,A2}. At the same time the antiferromagnetic exchange interaction opens a window
for formation of the Kondo insulator state.
It is naive to believe that in the Kondo insulator state the Hamiltonian of the Kondo lattice reduces to the one-particle Hamiltonian of the Anderson lattice, in which the Hubbard repulsion is not taken into account and the energies of an one-particle state of electrons located at the lattice sites lie near the Fermi energy \cite{LP}. The behavior of an electron liquid in the symmetric Anderson lattice is equivalent to that in the spin-$\frac{1}{2}$ Kondo lattice, which made it possible to understand the nature of the Kondo insulator state in the framework of the symmetric Anderson lattice  \cite{KA}. Interestingly, in the state of the Kondo insulator, the cell doubles \cite{KA}; therefore, the phase transition to the Kondo insulator state occurs similarly to the Peierls and Mott phase transitions \cite{K1}.

Using a mean field approach, we consider the solution of the spin-$\frac{1}{2}$  anisotropic Kondo lattice.
We shall show that local moments form a static $\mathbb{Z}_2$ field in which the band electrons move.
A configuration of the  $\mathbb{Z}_2$ field at which the energy of the system is minimal, corresponds to an antiferromagnetic order of local moments, the lattice cell doubles .The spectrum of quasi-particle excitations has the Majorana type, it is particle-hole symmetric in the Kondo insulator state. It should be noted the works that consider the Kondo lattice, taking into account the interaction proposed by Kitaev \cite{KS,M1,M2,M3}. These models make it possible to study the spin liquid with gapless spin excitations on the Majorana Fermi surface.
We declare that for an arbitrary dimension of the system, the Kondo insulator state  is realized on a lattice with a double cell. This result also follows from the solution of the symmetric Anderson lattice, obtained for different dimension \cite{KA}.

\section*{Model}

The Hamiltonian of the spin-$\frac{1}{2}$ Kondo lattice dimension D  ${\cal H}={\cal H}_0+{\cal H}_{K}$
includes two terms, the first of which is determined by energy of s-electrons, and  the second is determined by  the contact exchange interaction of these electrons with spins located at the lattice sites
\begin{eqnarray}
&&{\cal H}_0= -
\sum_{<i,j>}\sum_{\sigma=\uparrow,\downarrow}c^\dagger_{i,\sigma} c_{j,\sigma}, \nonumber \\
&&{\cal H}_K=
2J_z\sum_{j=1}^N s^z_jS^z_j+2J_x \sum_{j=1}^N(s^x_jS^x_j+s^y_jS^y_j),
\end{eqnarray}
where $c^\dagger_{j,\sigma}$ and $c_{j,\sigma}$ are the fermion operators determined on a lattice site $j$, $\sigma =\uparrow,\downarrow$ denotes the spin of electron, the hopping integral between the nearest-neighbor lattice sites is equal to one, the spin operators of s-electrons $s_j^{\alpha}=\frac{1}{2}c^\dagger_{j,\sigma}\sigma^{\alpha}_{\sigma,\sigma'}c_{j,\sigma'}$ are determined by the Pauli matrices
$\sigma^{\alpha}$ ($\alpha=x,y,z$), $\textbf{S}_j$ is the spin-$\frac{1}{2}$ operator defined on the lattice site $j$, the anisotropic exchange interaction $J_z,J_x >0$ creates a spin-flip electron scattering channel, we introduce also the following designation $u=J_z-J_x$ for description of anisotropy of the spin-exchange interaction in ${\cal H}_K $, N is the total number of lattice sites.

Let us define the spin-operators  $\textbf{S}_j$ using the $d^\dagger_{j,\sigma}$ and $d_{j,\sigma}$ fermion operators as
$S_j^{\alpha}=\frac{1}{2}d^\dagger_{j,\sigma}\sigma^{\alpha}_{\sigma,\sigma'}d_{j,\sigma'}$  with an additional constrain $n_j=n_{j,\uparrow}+n_{j,\downarrow}=1$, here $n_{j,\sigma}=d^\dagger_{j,\sigma}d_{j,\sigma}$ is the density operator. We use the $c_j$-operator for the conduction (hopping) electron and $d_j$-operator for the localized electron, originally coming from the s and d orbits.  We study an electron liquid state in the chain (1D) and on the square (2D) and cubic (3D) lattices at half-filled occupation.

\section*{The ground-state of the Kondo lattice}

\begin{figure}[tp]
     \centering{\leavevmode}
\begin{minipage}[h]{.32\linewidth}
\center{
\includegraphics[width=\linewidth]{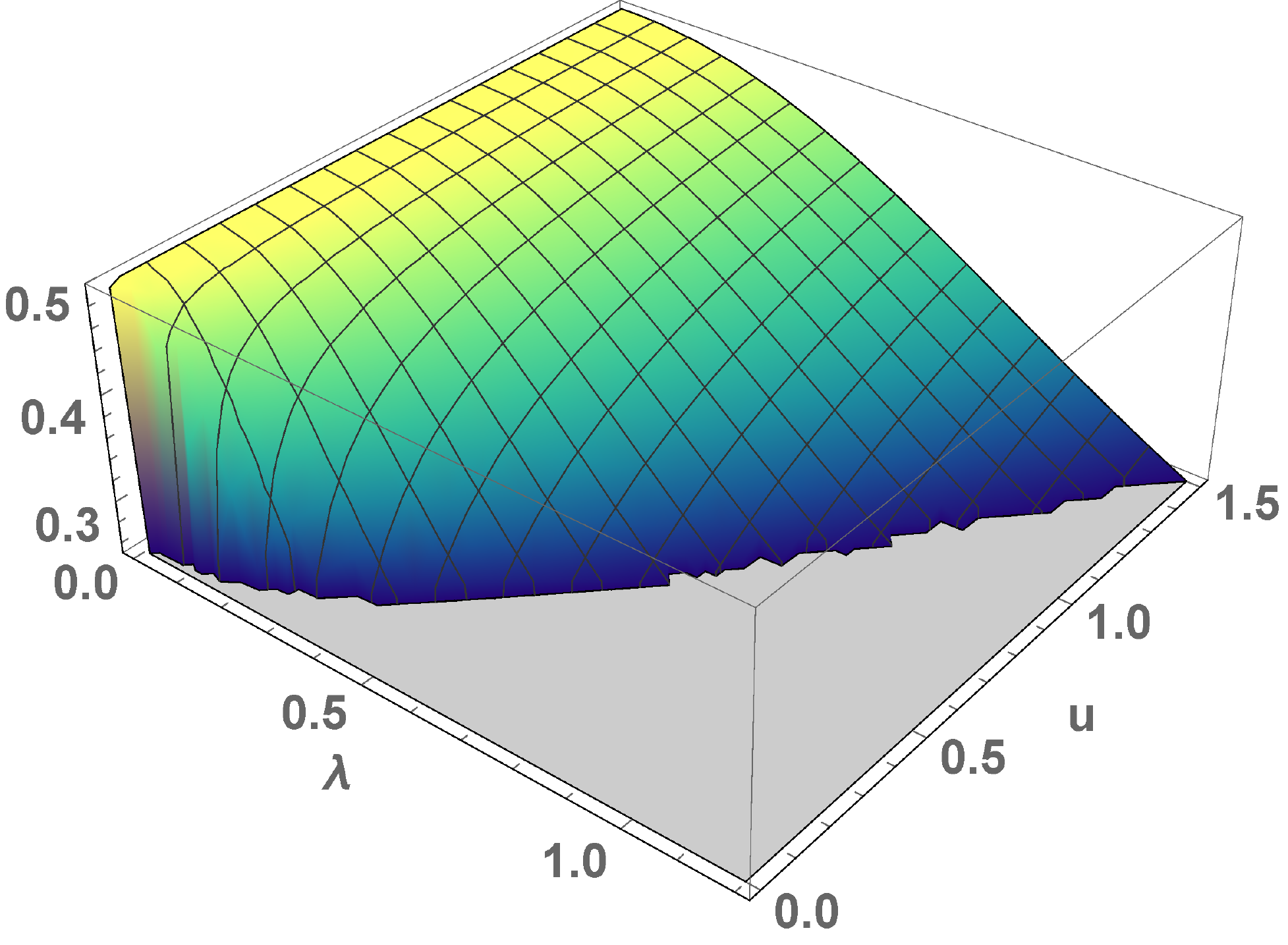} a)\\
                 }
   \end{minipage}
\begin{minipage}[h]{.32\linewidth}
\center{
\includegraphics[width=\linewidth]{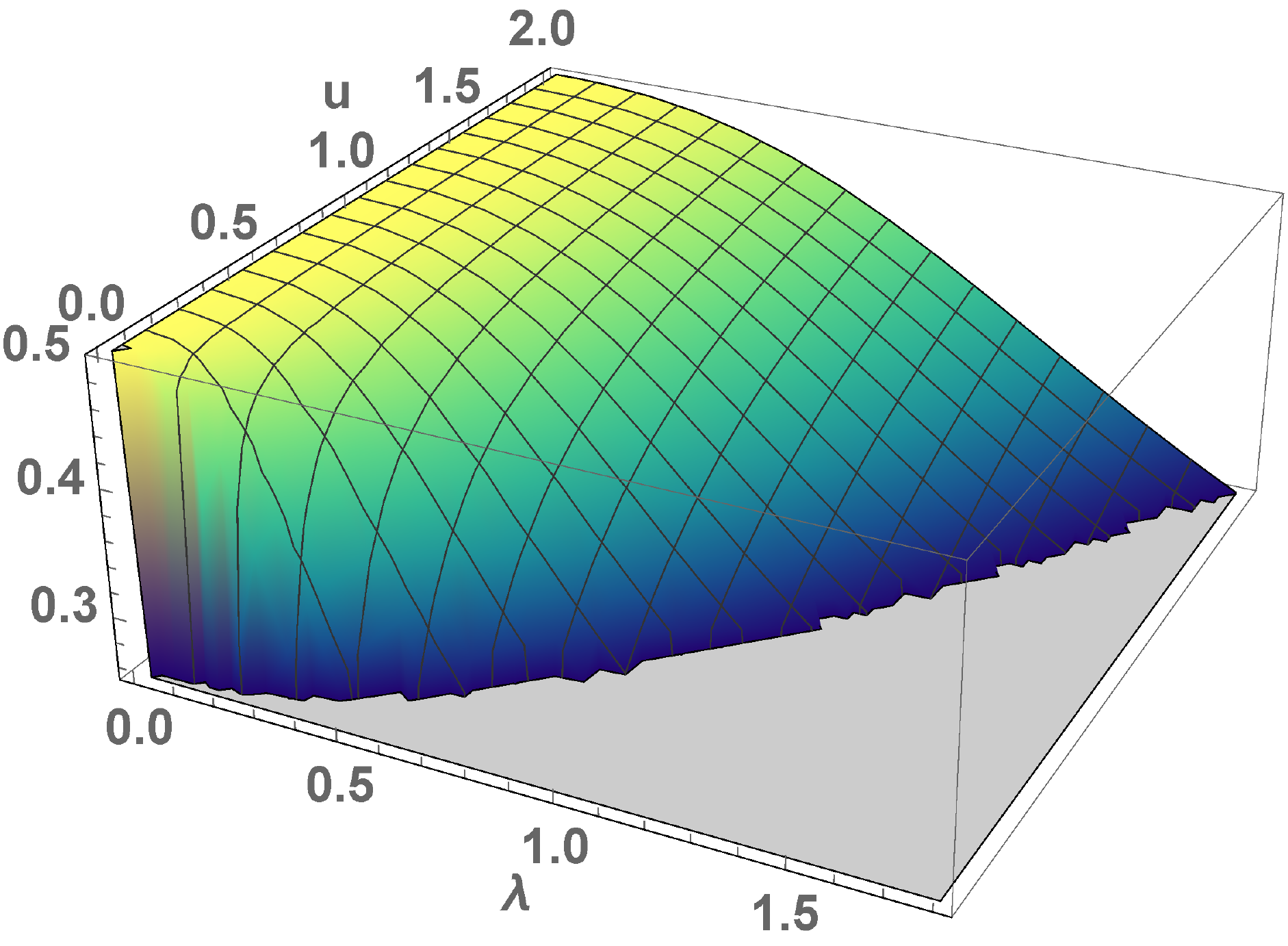} b)\\
                  }
   \end{minipage}
   \begin{minipage}[h]{.32\linewidth}
\center{
\includegraphics[width=\linewidth]{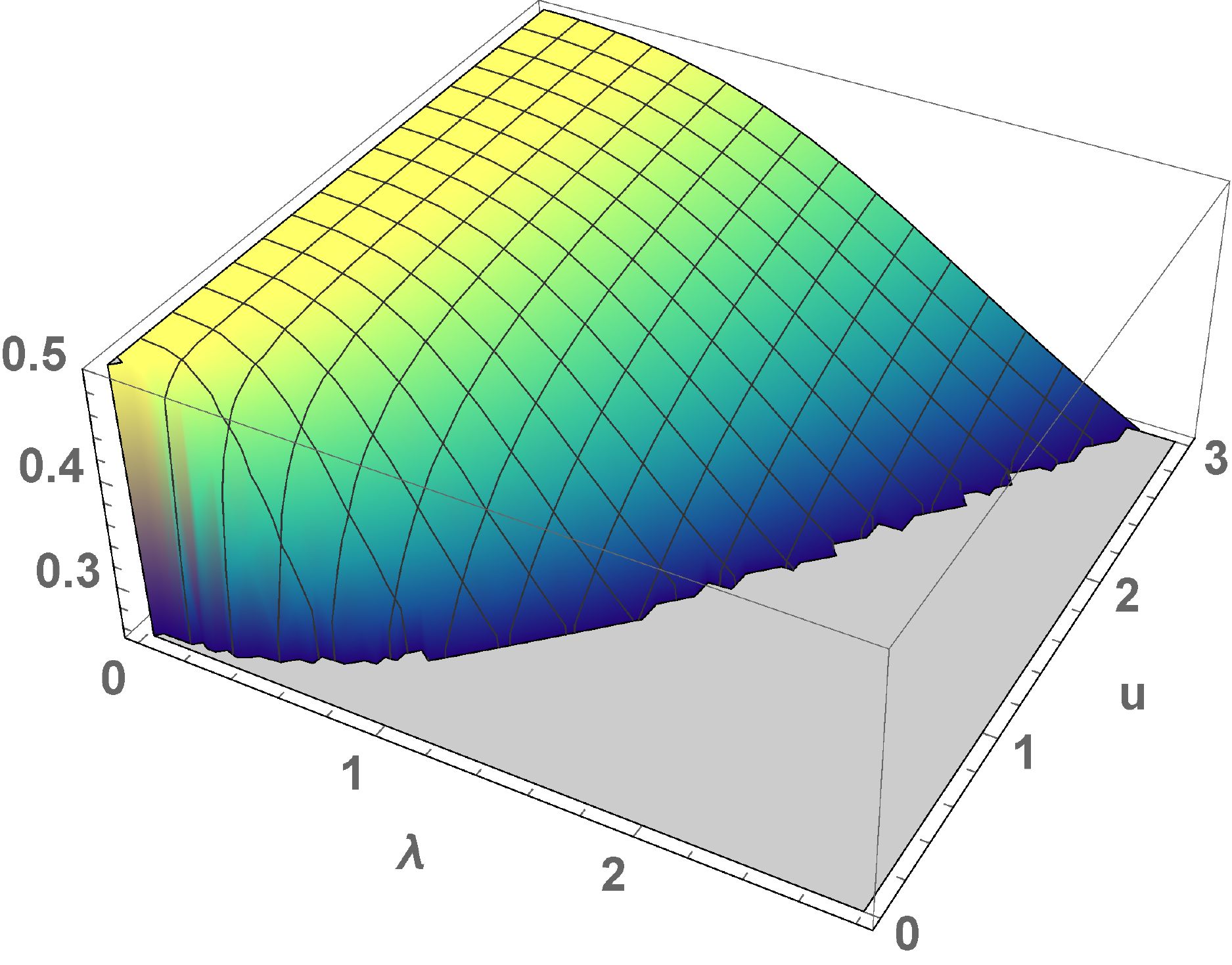} c)\\
                  }
   \end{minipage}
\caption{(Color online)
A local moment of $d$-electrons depending on $\lambda$ and $u$  calculated for the chain a), square b)  and cubic c) lattices, the region of $\lambda$, in which local moment changes from $\frac{1}{2}$ to $\frac{1}{4}$, is bounded  by  $\lambda_{max}$ for which  local moment is equal to $\frac{1}{4}$.
 }
\label{fig:1}
\end{figure}

Using the effective Hamiltonian (8) we calculate the energy of quasi-particle excitations, the spectrum includes four branches $\pm E_\gamma (\textbf{k)} $ ($\gamma =1,2$) symmetric about zero energy
\begin{equation}
E_\gamma(\textbf{k}) =\sqrt{\alpha(\textbf{k})+(-1)^\gamma\sqrt {\beta(\textbf{k)}}},
\label{eq:H5}
\end{equation}
where $\alpha (\textbf{k})=|\lambda|^2+\frac{1}{8}u^2+\frac{1}{2}|w(\textbf{k})|^2$,
$\beta (\textbf{k})=\frac{1}{4}|\lambda|^2 u^2 + \frac{1}{64}u^4+ |\lambda|^2|w(\textbf{k})|^2 - \frac{1}{8}u^2|w(\textbf{k})|^2 +\frac{1}{4} |w(\textbf{k})|^4 $,
 $w(\textbf{k})=\sum^D(1+\exp(i k_\alpha))$, $\textbf{k}=(k_x,k_y,k_z)$ is the wave vector.

The amplitude of the $\lambda$-field determines the phase state of electron liquid, its value corresponds to minimum of the action (see section Methods).
Nontrivial solution for $\lambda$ leads to open a gap in the electron spectrum $\Delta$. The value of the gap has an universal behavior as a function of $ \lambda$, it value does not depend on the dimension of the lattice \cite{KA},
at $u>\sqrt{2} \lambda$ $\Delta =\sqrt{4 |\lambda|^2+ u^2/2-u\sqrt{4|\lambda|^2+u^2/4}}$ and $u<\sqrt{2} \lambda$  $\Delta =\sqrt{8+4 |\lambda|^2+ u^2/2-\sqrt{(u^2/2-8)^2+4|\lambda|^2(16+u^2)}}$. In the case $u>0$ the gap in the spectrum opens only when $\lambda\neq 0$.

The solution for $\lambda$ corresponds to minimal action (9), in the saddle point approximation a self-consistent equation has the following form at $T=0K$
\begin{equation}
2\frac{\lambda}{J_x}=\frac{1}{N}\sum_{\gamma=1,2}\int d \textbf{k} \frac {\partial E_\gamma(\textbf{k})}{\partial \lambda}.
\label{eq:H6}
\end{equation}
$\lambda$  is the solution of Eq (3), for given $J_x$ its value is determined by $J_z$, depends on dimension of the lattice.
Due to symmetric spectrum of quasi-particle excitations (2), the density of $d-$electrons $n=\sum_{\textbf{k},\sigma}n_{\textbf{k},\sigma}$
is equal to 1. Such the density of states of d-electrons with spin $\sigma$ and $-\sigma$ are equal to $n_{\sigma}=1/2+\delta n, n_{-\sigma}=1/2-\delta n$
\begin{equation}
 \delta n =\frac{u}{4 N}\sum_{\textbf{k}}\frac{1}{E_2(\textbf{k})+E_1(\textbf{k})} \left(\frac{|w(\textbf{k})|^2}{E_1(\textbf{k}) E_2(\textbf{k})}+1\right).
\label{eq:H5}
\end{equation}
z-projection of a local moment equal to $S_{loc}=\frac{1}{2}(n_{\sigma}-n_{-\sigma})=\delta n$, is equal to zero in the $\lambda$ infinity limit.

A reasonable limitation would be to consider small values of $\lambda<\lambda_{max}$, at which the local moment is decreased from $\frac{1}{2}$ to $\frac{1}{4}$, without  taking into account large values of $\lambda$, which correspond to a smaller value of the local moment,
where $\lambda_{max}$ corresponds to $d$-states with $S_{loc}=\frac{1}{4}$.
Numerical calculations of $S_{loc}$ depending on $\lambda$ (where $\lambda<\lambda_{max}$) and $u$, obtained for different lattice dimension, are shown in Figs 1.
Consider a lattice with local moments, that is, for $\lambda$ less than $\lambda_{max}$ (see Figs 1), such for $u = 0.001$, $\lambda_{max} = 0.0316 (1D), 0.045(2D), 0.052 (3D)$; $u = 0.01$  $\lambda_{max} =0.0994 (1D),  0.135 (2D),  0.1635 (3D)$;
$u=0.1$, $ \lambda_{max}=0.307 (1D), 0.4185 (2D),  0.51 (3D)$; $u=0.5$, $\lambda_{max}=0.67 (1D), 0.903 (2D), 1.099 (3D)$;
$u=1$,  $\lambda_{max}=0.96 (1D), 1.264 (2D), 1.53 (3D)$. As expected, the local moment at a lattice site is more than $\frac{1}{4}$ for $\lambda^2<u$,
as the lattice dimension increases, this condition becomes less stringent. In general, the mean field approach becomes more plausible in higher dimensions (ultimately $D\to \infty$).
Unfortunately, this approach does not allow us to study the behavior of an electron liquid in the symmetric Kondo lattice model. The problems are due to the fact that the representation for spin operators in terms of electron operators does not adequately define the local spin for the isotropic exchange interaction.
The $\lambda$-field breaks the conservation the number of $d$- and $s-$  electrons, due to s-d hybridization only total number of electrons is conserved. As a result, a large Fermi surface defines the density of electrons.

\begin{figure}[tp]
     \centering{\leavevmode}
\begin{minipage}[h]{.35\linewidth}
\center{
\includegraphics[width=\linewidth]{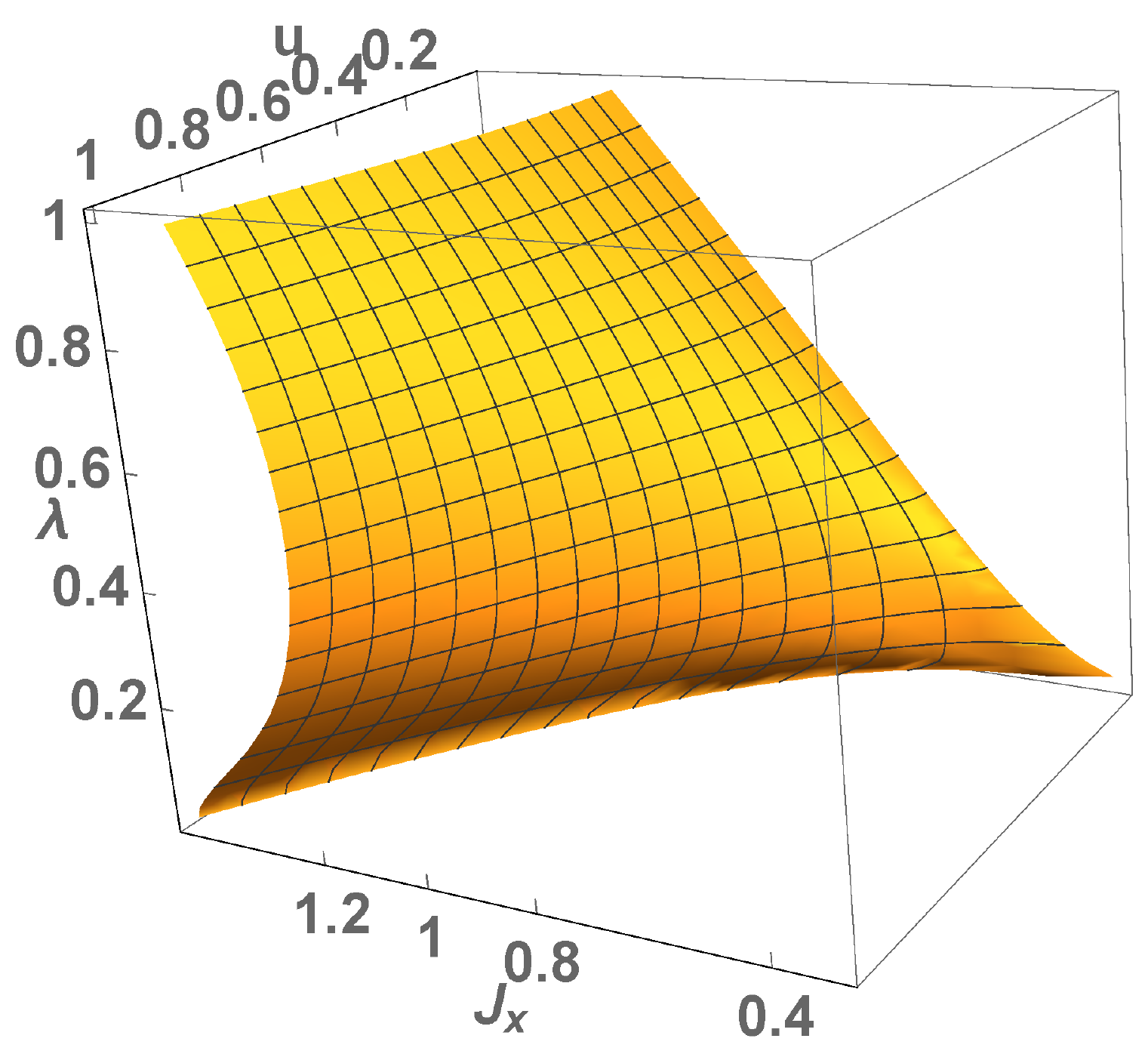} a)\\
                 }
   \end{minipage}
\begin{minipage}[h]{.25\linewidth}
\center{
\includegraphics[width=\linewidth]{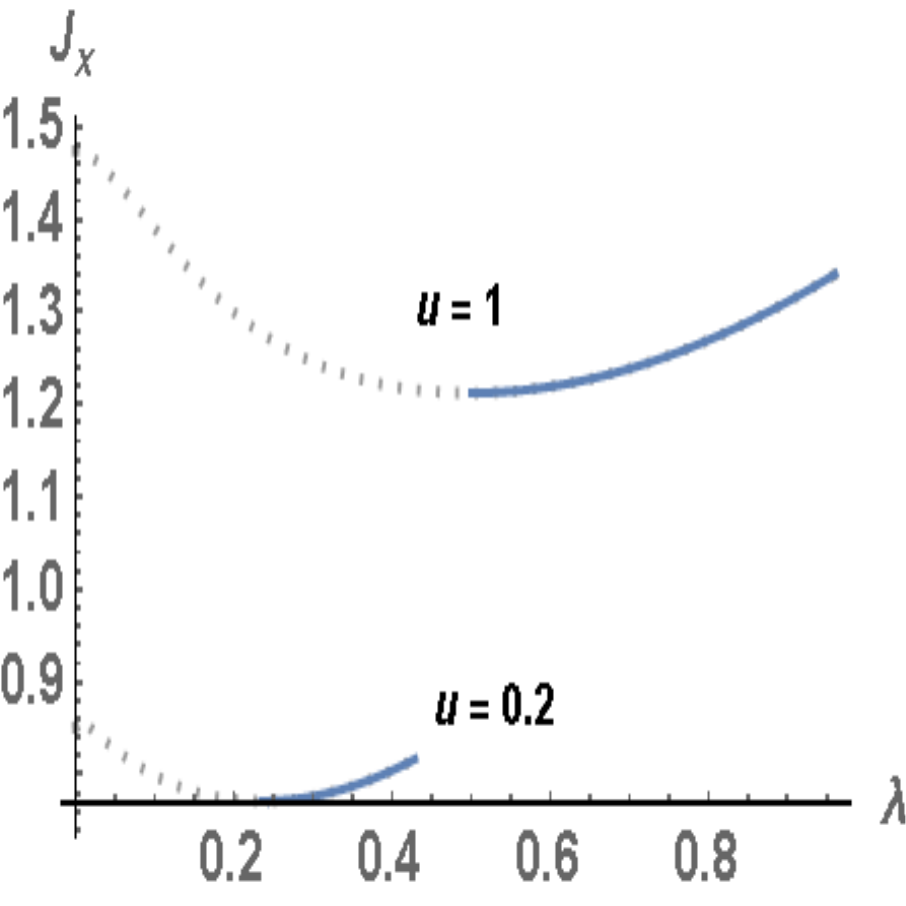} b)\\
                  }
   \end{minipage}
   \begin{minipage}[h]{.35\linewidth}
\center{
\includegraphics[width=\linewidth]{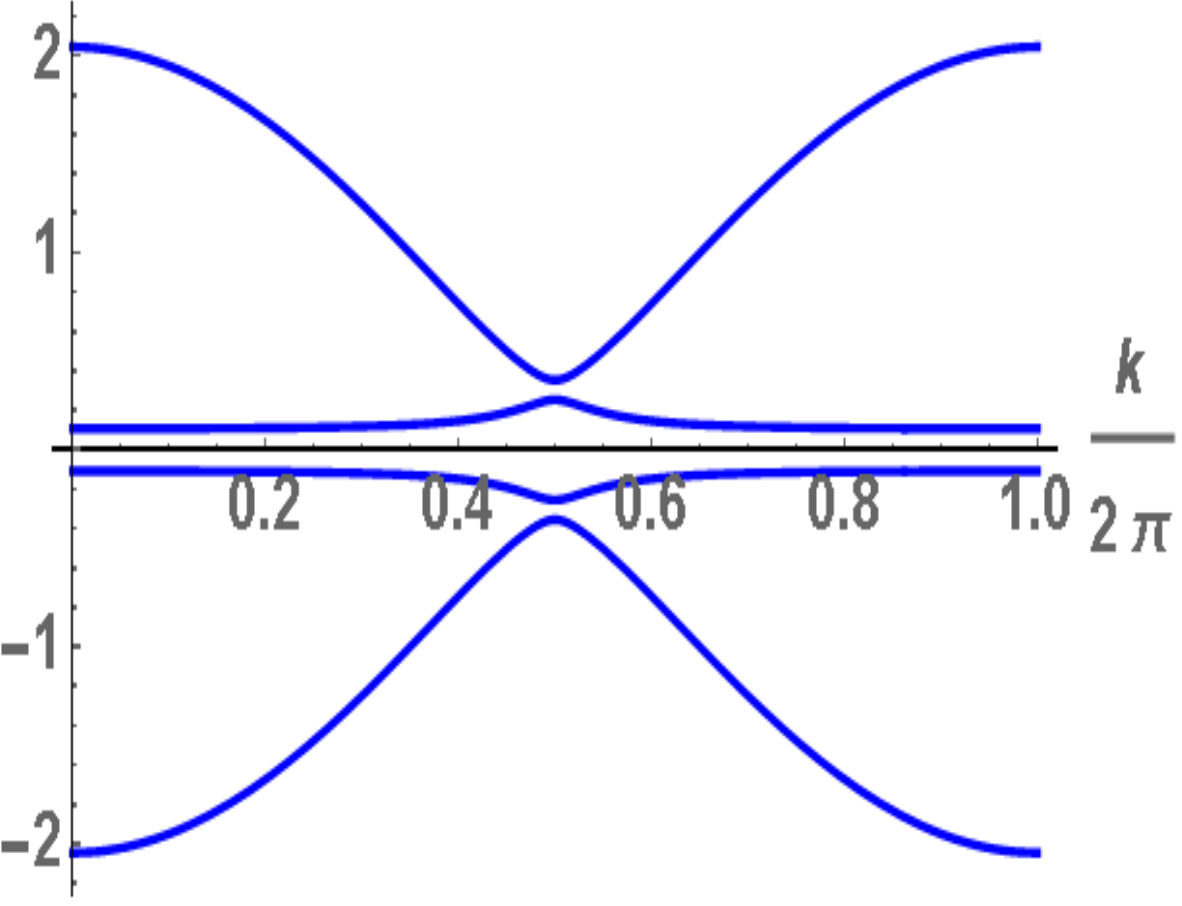} c)\\
                  }
   \end{minipage}
\caption{(Color online)
$\lambda$ as a function of $J_x$ and $u$  calculated for the chain a), calculations of $J_x$ as a function of $\lambda$ in  weak at $u=0.2$ and strong at $u=1$ anisotropy of the exchange interaction b) (dotted and blue lines correspond to instable and stable states of electron liquid). A spectrum  of the Kondo chain c) as  a function of the wave vector calculated at $u=0.2$, $\lambda=0.3$ and $J_x=0.776$.
 }
\label{fig:2}
\end{figure}

\begin{figure}[tp]
     \centering{\leavevmode}
\begin{minipage}[h]{.3\linewidth}
\center{
\includegraphics[width=\linewidth]{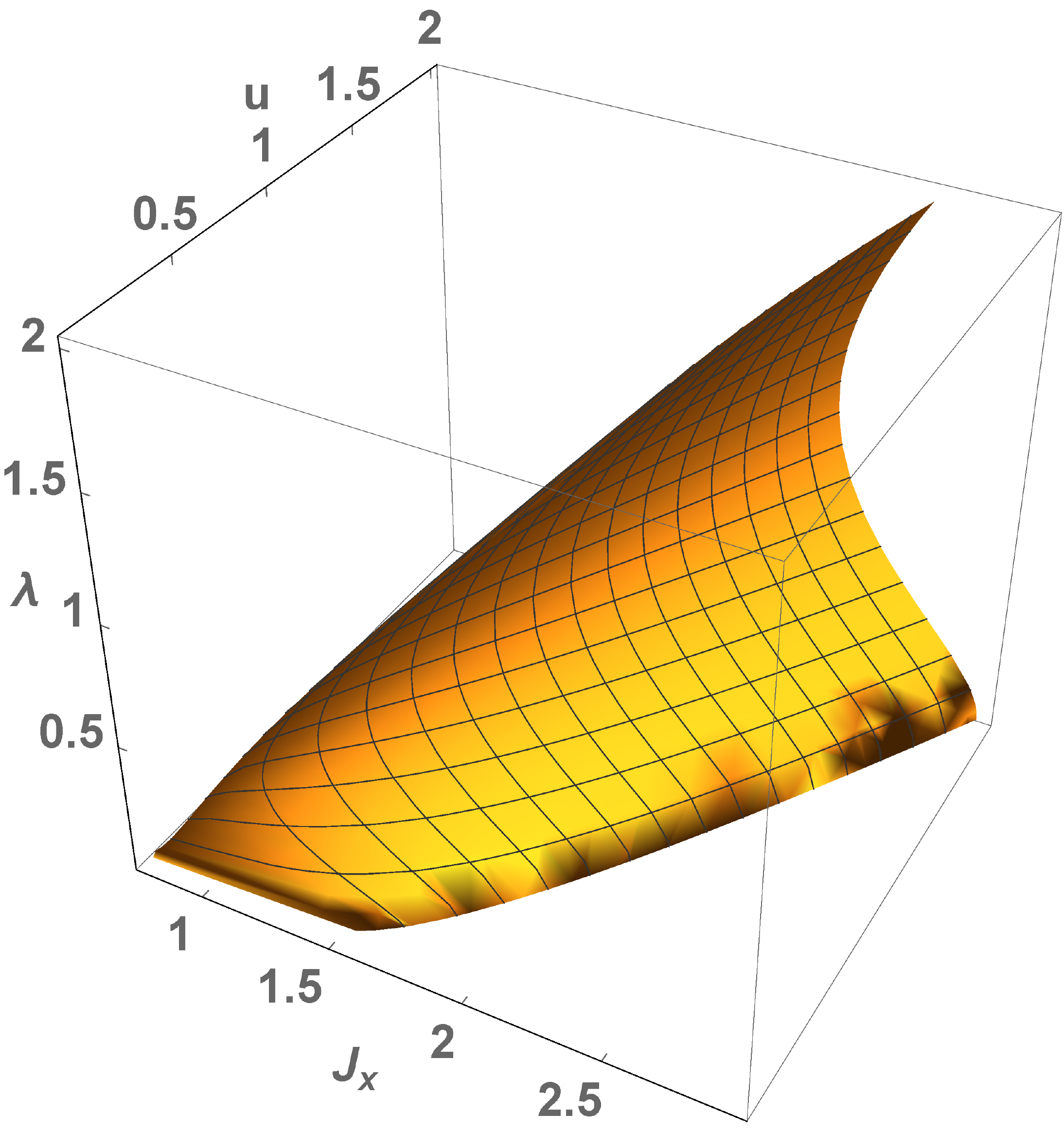} a)\\
                 }
   \end{minipage}
\begin{minipage}[h]{.28\linewidth}
\center{
\includegraphics[width=\linewidth]{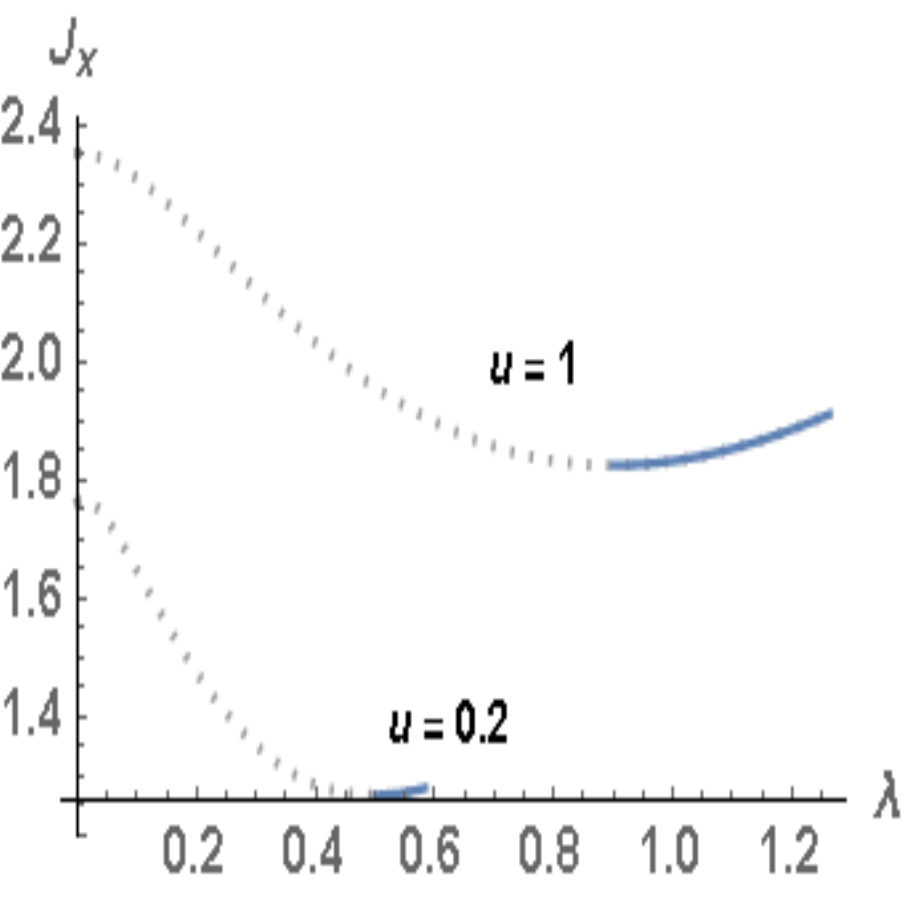} b)\\
                  }
   \end{minipage}
   \begin{minipage}[h]{.4\linewidth}
\center{
\includegraphics[width=\linewidth]{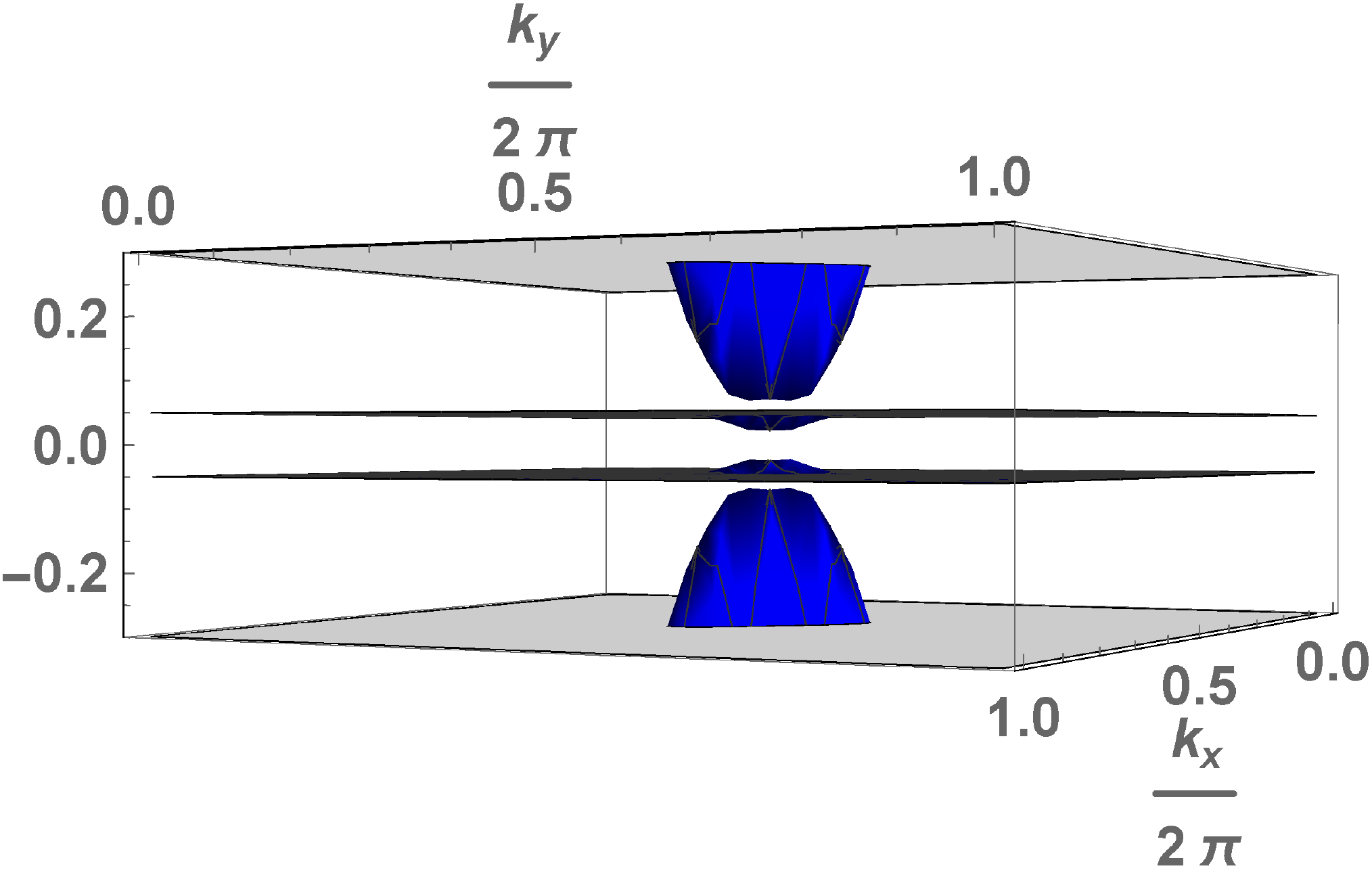} c)\\
                  }
   \end{minipage}
\caption{(Color online)
$\lambda$ as a function of $J_x$ and $u$  calculated for the square lattice a), $J_x$ as s function of $\lambda$ b), a low energy spectrum as a function of the wave vector  calculated at $u=0.2$, $\lambda =0.5$ and $J_x=1.268$.
 }
\label{fig:3}
\end{figure}

\begin{figure}[tp]
     \centering{\leavevmode}
\begin{minipage}[h]{.32\linewidth}
\center{\includegraphics[width=\linewidth]{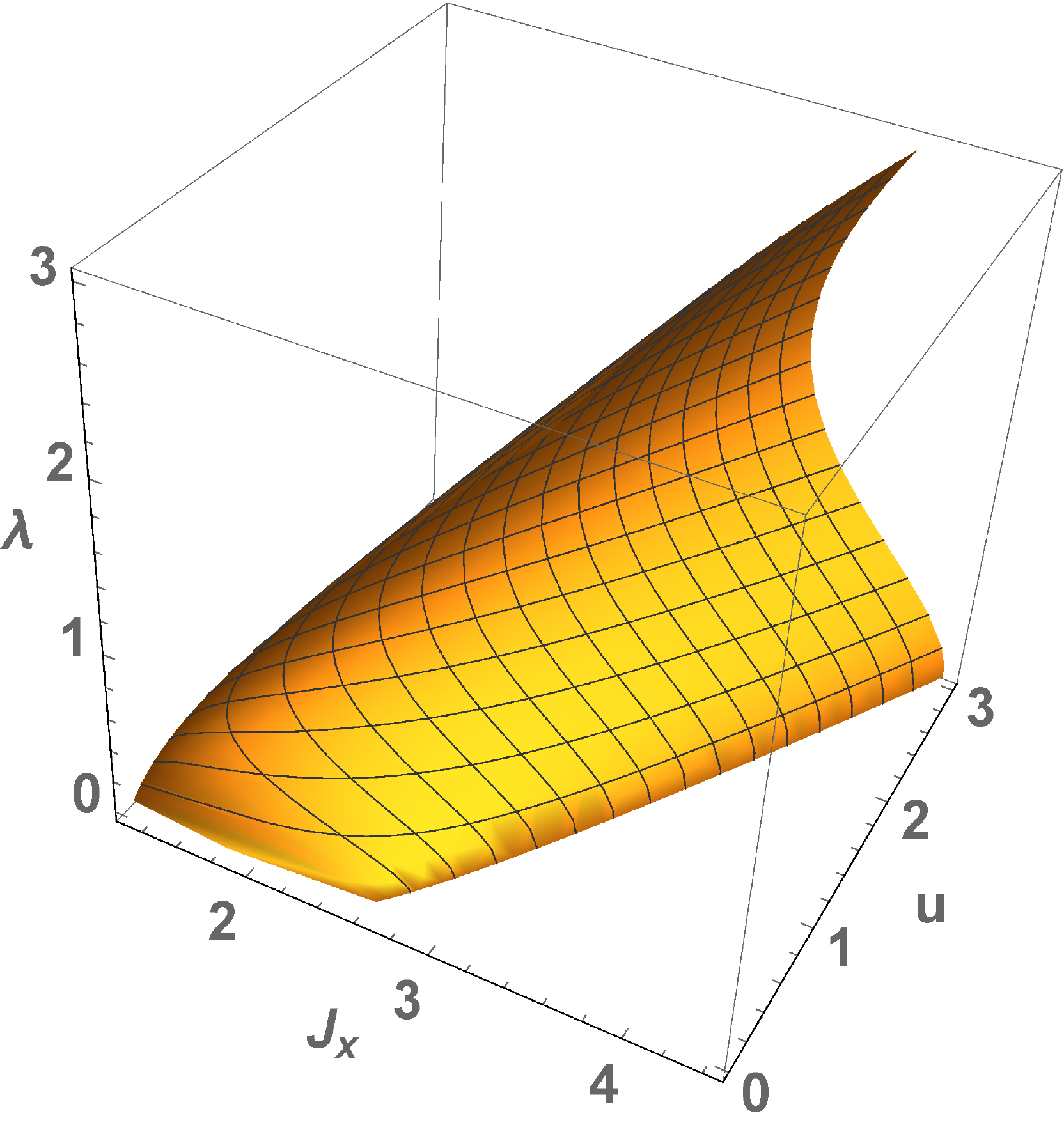} a)\\
                 }
   \end{minipage}
\begin{minipage}[h]{.4\linewidth}
\center{
\includegraphics[width=\linewidth]{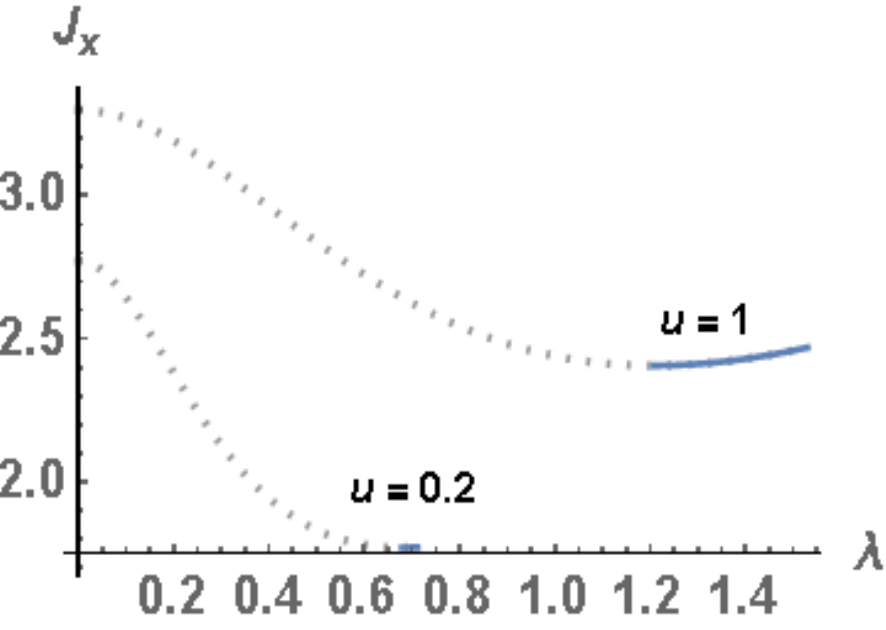} b)\\
                  }
   \end{minipage}
\caption{(Color online)
$\lambda$ as a function of $J_x$ and $u$ calculated for the cubic lattice a), $J_x$ as a function of $\lambda$ in  weak at $u=0.2$ and strong at $u=1$ anisotropy of the exchange interaction b).
 }
\label{fig:4}
\end{figure}

As we noted above in the electron spectrum  the gap opens at $\lambda\neq 0 $, its value is also determined by $J_x$ and $J_z$ or $u$.
Using Eq (3) we numerically calculate $\lambda$ as function of $J_x$ and $u$, the results of the calculations for different dimension of the lattice are presented in Figs 2a,3a and 4a. To illustrate, let us analyze the results obtained in the case of weak and strong anisotropy of the exchange interaction at $u=0.2$ and $u=1$ for the chain Figs 2, square and cubic lattices Figs 3,4.
From the numerical results it follows that for given $u$ and $J_x$  there are two non-trivial solutions of $\lambda$. We also take into account that $\lambda<\lambda_{max}$. In Figs 2b, 3b  and 4b the value of $J_x$ is shown as a function of $\lambda$ for fixed values of $u$. The minimum of ground state energy corresponds to maximal value of $\lambda$, therefore the instable branch of the solution is marked by dotted line, a stable branch is marked by blue line.

In the Kondo chain a gapped state of electron liquid is realized for arbitrary values of $u$, such at $u \to 0$ solution for $J_x$ also $J_x\to 0$ (see in Figs 1a),2a). In the square  lattice a minimal value of $u_{min}$ is equal to $0.04$ which corresponds to $\lambda=0.267$ and $J_x=1.22$, other words the gapped state is realized at $u>0.04$ and $J_x>1.22$ (see in Figs 1b), 3a). In cubic lattice a minimal value of $u$ is equal to $0.13$, at which corresponds to $\lambda=0.58$ and $J_x=1.68$ (see in Figs 1c),4a). A fixed value of $u$ corresponds to nontrivial solution of $\lambda$, such a quantum phase transition in gapped state is a first order phase transition. The electron spectra  (in 1D) and its low energy part (in 2D) are shown  in Fig.2c and Fig.3c, the calculations are given for the case of a weak anisotropy of the exchange interaction at $u=0.2$. There are two gaps in the quasi-particle spectrum: the low energy one is insulating gap $\Delta$, the high energy one separates the branches of s- and d-electrons.

\section*{Methods}
We introduce the operator $\chi_{j}= c^\dagger_{j, \uparrow}d_{j,\uparrow}+c^\dagger_{j,\downarrow}d_{j,\downarrow}$  \cite{CA} and redefine the term
${\cal H}_{K}$ in the following form
\begin{eqnarray}
&& {\cal H}_{K}= 2u\sum_{j=1}^N s^z_jS^z_j-J_x\sum_{j}\chi^\dagger_j \chi_j -\frac{1}{2}J_x\sum_j n_j m_j,
\end{eqnarray}
here $m_j=\sum_\sigma  c^\dagger_{j, \sigma} c_{j, \sigma}$ is the density operator.
Let us consider the solution of the problem for $ J_z> J_x$ and $u>0$. The $zz$-exchange interaction in (5) determines the energies of the states  of $d-$electrons located at the lattice sites, one particle state has the energy $-\frac{u}{2}$, the energy of two particle state is equal to zero.
Two $d$-electrons located at site $j$ have the energies: $-\frac{u}{2}$ for the first  and $\frac{u}{2}$ for the second. This symmetric about the Fermi energy arrangement of levels realizes the constrain $n_j=1$ \cite{KA}.
In this case the state with $n_j=1$ is realized, as it takes place in the symmetric Anderson lattice \cite{KA}.
The Hubbard-Stratonovich transformation reduces the exchange interaction to a non-interacting ones in a stochastic external $\lambda$-field, which determines by the interaction strength. Taking into account the action ${S}_{0}$ (the action with respect to the free Hamiltonian ${\cal H}_0$), we redefine the interaction term as follows $S=S_{0}+\frac{1}{J_x}\sum_{j}\lambda^*_j\lambda_j+\sum_{j}(\lambda^*_j\chi_{j}+\lambda_j\chi^\dagger_j)$.
The canonical functional is determined by the action
\begin{eqnarray}
&& S=\frac{1}{J_x}\sum_{j} \lambda^*_j \lambda_{j}+
\int_0^\beta d\tau \sum_\textbf{k}\Psi_\textbf{k}^\dagger (\tau)[\partial_\tau  + {\cal H}_{eff}(\textbf{k})]\Psi_\textbf{k} (\tau),
\end{eqnarray}
where  $\Psi_\textbf{k} (\tau)$ is  the wave function, $\textbf{k}$ is the wave vector of an electron. The effective Hamiltonian ${\cal H}_{eff}(\textbf{k})$ determines low-energy excitations of electrons at half filling occupation

We expect that $ \lambda_{j} $  did not depend on $ \tau $, since in an electron liquid state, the translation invariance is conserved. $\lambda$-field leads to the on-site hybridization between s- and d-states of electrons.
Let us consider the equations for the one-particle wave functions $\psi(\textbf{j},\sigma)c^\dagger_{\textbf{j},\sigma}+\phi (\textbf{j},\sigma)d^\dagger_{\textbf{j},\sigma}$ with energy $\epsilon$:
\begin{eqnarray}
&&\epsilon \psi(\textbf{j},\sigma) +\lambda \phi(\textbf{j},\sigma)+\sum_{\textbf{1}}\psi(\textbf{j+1},\sigma)=0,
 \left(\epsilon -\frac{u_\textbf{j}}{2}\right) \phi(\textbf{j},\sigma) +\lambda^* \psi(\textbf{j},\sigma)=0,\nonumber\\
&&\epsilon \psi(\textbf{j},-\sigma) +\lambda \phi(\textbf{j},-\sigma)+\sum_{\textbf{1}}\psi(\textbf{j+1},-\sigma)=0,
\left(\epsilon +\frac{u_\textbf{j}}{2}\right)\phi(\textbf{j},-\sigma) +\lambda^* \psi(\textbf{j},-\sigma)=0,
\end{eqnarray}
where sums  over the nearest lattice sites and $\lambda_j$ does not depend on $j$. The variables  $u_j=\pm u$  are identified with a static $\mathbb{Z}_2$ field determined on the lattice sites, the band electrons move in this static field. Note that the energy of $d$-electron is not defined by its spin, because only the energies of one- and two-particle $d$-states on the lattice site are fixed, which are equal to $-\frac{u}{2}$ and $0$, respectively.
Detailed numerical analysis shows, that an uniform sector with $u_\textbf{j} =u$ and $u_\textbf{j+1}=-u$ for all variables is the ground state of an electron liquid described by Eqs (7) \cite{KA,Lieb,Kitaev,KS}.

Let us consider the antiferromagnetic state, when $\sum_{{j}}<c^\dagger_{j,\uparrow}c_{j,\uparrow}>= \sum_{j}<c^\dagger_{j,\downarrow} c_{j,\downarrow}>$.
Due to the translational invariance of the lattice, the equations for the wave function are solved analytically for the uniform configuration of a static $\mathbb{Z}_2$ field. The effective Hamiltonian ${\cal H}_{eff}(\textbf{k})$, which corresponds to uniform configuration with a double lattice cell, has the following form
\begin{equation}
{\cal H}_{eff}(\textbf{k}) = \left(
\begin{array}{cccc}
0 & \lambda& w(\textbf{k}) & 0\\
\lambda^* & \frac{u}{2} & 0& 0\\
w^*(\textbf{k}) & 0& 0& \lambda \\
0 & 0&\lambda^* &- \frac{u}{2}
\end{array}
\right).
\end{equation}

We can  integrate out fermions to obtain  the following action $S$ per an atom
\begin{eqnarray}
S(\lambda)=-\frac{{T}}{{N}}\sum_{\textbf{k}}\sum_n \sum_{\gamma=1}^{4} \ln [-i \omega_n+\varepsilon_\gamma(\textbf{k})]
+\frac{|\lambda|^2}{J_x},
 \label{A2}
\end{eqnarray}
where $\omega_n =T(2n+1)\pi$ are the Matsubara frequencies,  four quasiparticle excitations $\varepsilon_\gamma(\textbf{k})$ ($\gamma =1,...,4$) determine the electron states in the  $\lambda$-field. In the saddle point approximation the canonical functional will be dominated by the minimal action $S$ (9), that the spectrum of the quasi-particle excitations is symmetric about zero energy, it has the Majorana type. The opening of the gap occurs due to the doubling of the cell in the insulator state.

\section*{Conclusion}

We studied the behavior of electron liquid in the spin-$\frac{1}{2}$ anisotropic Kondo lattice at half-filling for different dimension.
The calculation results are valid for an anisotropic Kondo lattice with $J_z >J_x $, since the constrain $n_j=1$  and local moments at the lattice sites are realized only in the case of an  anisotropic exchange interaction.
It is shown that band electrons move in  a static $\mathbb{Z}_2$ field of local moments, the uniform configuration of the $\mathbb{Z}_2$ field corresponds to ground state of electron liquid and leads to formation of a lattice with a double cell. In the insulator state the fermion spectrum is electron-hole symmetric as it takes place for the Majorana spectrum.
The gapped state is formed at finite values of the $J_z$ and $J_x$ exchange integrals. The results of calculations are valid in the case a weak anisotropy of the exchange interaction, that alow to made conclusion that the behavior of electron liquid in an isotropic Kondo lattice is similar.

\section*{Acknowledgments}
The second month of the war between Ukraine and Russia is underway, the author thanks the Ukrainian army and the Ukrainian people, who courageously defend Ukraine and Kyiv.

\section*{Author contributions statement}
I.K. is an author of the manuscript
\section*{Additional information}
The author declares no competing financial interests. \\
\section*{Availability of Data and Materials}
All data generated or analysed during this study are included in this published article.\\
Correspondence and requests for materials should be addressed to I.N.K.
\end{document}